\documentclass[preprint,12pt]{elsarticle}

\usepackage{amssymb}

\usepackage{color}
\usepackage{graphicx}
\usepackage{graphics}
\usepackage{amsmath}
\usepackage{amssymb}
\usepackage{amsfonts}
\usepackage{mathrsfs}
\usepackage[latin1]{inputenc}

\usepackage{mathrsfs}
\usepackage{stmaryrd}
\usepackage{latexsym}
\usepackage{bm}

\newcommand\be{\textbf{\emph{e}}}

\newcommand\f{\textbf{\emph{f}}}

\newcommand\bE{\textbf{\emph{E}}}
\newcommand\bD{\textbf{\emph{D}}}

\newcommand\F{\textbf{F}}

\newcommand\B{\textbf{B}}

\newcommand\I{\textbf{I}}
\newcommand\T{\textbf{T}}

\renewcommand\d\delta
\newcommand\D\Delta
\newcommand\e{\varepsilon}

\newcommand\scrD{\mathscr{D}}

\newcommand\beq{\begin{equation}}
\newcommand\eeq{\end{equation}}

\journal{}

\begin{document}

\begin{frontmatter}



\title{Compression-induced failure of electro-active polymeric thin films}


\author{D. De Tommasi, G. Puglisi, G. Zurlo}

\address{d.detommasi@poliba.it}

\begin{abstract}
The insurgence of compression induces wrinkling in actuation devices based on EAPs thin films leading to a sudden decrease of performances up to failure. Based on the classical tension field theory for thin elastic membranes
(e.g. \cite{Pipkin1}), we provide a general framework for the analysis of the
insurgence of in-plane compression in membranes of electroactive polymers
(EAPs).  Our main result is the deduction of a (voltage-dependent) domain in the stretch space which represents tensile configurations. Under the assumption of Mooney-Rivlin materials, we obtain that for  growing values of the applied voltage the domain contracts, vanishing at a critical voltage above which the polymer is wrinkled for any stretch configuration.  Our approach can be easily implemented in numerical simulations for more complex material behaviors and provides a tool for the analysis of compression instability as a function of the elastic moduli.  \end{abstract}


\end{frontmatter}



\noindent The growing interest in electroactive polymers as actuator devices,
ranging from medical, biological, robotic, and energy harvesters, results from
their qualities such as lightweight, small size, low-cost, flexibility, fast
response \cite{KP, KZS, CRK}. A typical  device consists of a thin sheet of
electroactive polymer sandwiched between two compliant electrodes. The simple
mechanism of actuation releases on an electromechanical coupling of the Coulomb
forces acting between the electrodes and the elastic forces inside the
layer. The electrostatic forces acting on the sheet faces induce a transversal extension that is
used as a mean of actuation.

In this paper we are mainly concerned with compression induced instability
phenomena of thin polymeric electroactive films. The technological interest on
these phenomena is due to the observation that, aside of purely electric
breakage, typical failure mechanisms of EAPs technological applications are induced by sudden loss of equilibrium.
This instability is due to the thinness of the layer and the consequent
inability of sustaining compressive stress. As a result, the importance of prestraining in improving the actuation properties has been evidenced in several papers, (e.g. \cite{K, KP, PD, PKP, SGS}), where the authors describe, as predicted by our theory,  the existence of optimal prestretch
values. A theoretical analysis of the
insurgence of deformation localization in a variational framework was recently
proposed in \cite{DPSZ} and \cite{ZHS}, and in \cite{GP} where the role of damage and dissipation
were also analyzed.

Here, we obtain explicit analytical results for Mooney-Rivlin incompressible materials, evaluating
the insurgence of compressive instability for a generic membrane . While analytical and
numerical results about this phenomenon were already obtained in other articles
(see \cite{DPSZ}, \cite{PD} and references therein) under restrictive assumptions on the homogeneity of deformation and on the device geometry, a general analytical approach to this topic is still not available
 up to the knowledge of the authors.

Our results take inspiration on the tension field theory for elastic membranes
(\cite{Pipkin1}, \cite{SteigPip}, \cite{Steig1}). The main ingredient of the
theory is the existence of a \emph{natural width}, assigning  a threshold  of one of the in-plane stretches as a function of the other one. This threshold separates compressed and tensile states. Accordingly, in the quoted papers it is shown the possibility of decomposing the stretch space into a domain characterized by positive principal stresses (\emph{tensile configurations}), a region where one stress is positive and the other is negative (\emph{wrinkled configurations}), and the remaining region where both stresses are negative.

Here we extend these results to the analysis of electroactivated membranes. As
we show, for sufficiently high values of the assigned voltage, the tensile region
reduces to an ``island", that we can analytically describe and that shrinks as
the voltage is increased. We then deduce the existence of a loading threshold
(critical voltage), such that for larger value of the electric load no tensile
configuration is possible. The amplitude of the safe stretch region and the
critical threshold strongly depend on the constitutive properties of the
material: ``stiffer" materials are safer.

We point out that our approach can be extended to general constitutive hypotheses. Moreover, our paper does not assume homogeneous deformations and delivers a framework to describe general boundary value problems for thin films of electroactivated materials.
We believe that the proposed approach will be useful not only to clearly
understand the insurgence of wrinkled configurations and the possible
disappearance of stable equilibrium states, but also because it delivers an instrument to study the behavior regarding the compression instability as a function of material moduli. This aspect is fundamental in the field of the design of new electroactive materials, a very active area of scientific and technological research.

To show the ability of the model of putting the subject in the right perspective
and clearly describe the physical ingredients of the phenomenon, at the end of
the paper we deliver two specific applications to simple boundary value
problems amenable of fully analytical results.



\section{Preliminary notions}

We here collect the main equations for a
continuum body under electromechanical loading.
 We refer the reader to \cite{BDO} and to
the references therein for details.

Let
 $\f$ be the deformation carrying the continuum body $\cal B$ (reference configuration) to the current configuration $\cal B'=\f (\cal B)$. We denote by
$\F=\nabla\f$ the deformation gradient, by
$\B=\F\F^T$ the
left Cauchy-Green tensor, and by ${\be}_i$ and $\lambda^2_i$ the eigenvectors and eigenvalues of $\B$,
where the $\lambda_i$ are the
principal stretches. $\bD$ and $\bE$ are the electric displacement and the electric field in the current configuration $\cal B'$,
respectively. For a linear, homogeneous and isotropic dielectric materials
$\bD = \e\bE $ where $\e=\e_o\e_d$ with
$\e_0$ the permittivity of free space and $\e_d$ the dielectric
constant of the material.

The (current) Cauchy
stress tensor $\T$ in the case of electromechanical body can be decomposed as the sum of the elastic stress tensor
$\T^{{el}}$ and of the electric Maxwell stress tensor $\T^{\tiny \mbox{\it M}}$:
$$\T=\T^{el}+\T^{\tiny \mbox{\it M}}.$$

We consider an incompressible, isotropic, elastic materials, for which $\det \F=\lambda_1\lambda_2\lambda_3=1$ and the elastic stress can be represented as a function of $\B$ (e.g. \cite{TN} Eq.(49.5)) as follows:
\begin{equation}
\begin{array}{ccl}\label{Cauchy}
\T^{el} & = &
-\pi\I+\beta_1(\lambda_1,\lambda_2)\B+\beta_2(\lambda_1,\lambda_2)\B^{-1}\\
T^{el}_{ij} & = &
-\pi\d_{ij}+\beta_1(\lambda_1,\lambda_2)B_{ij}+\beta_2(\lambda_1,\lambda_2)B_{ij}^{-1},
\end{array}
\end{equation}
where $\beta_1$ and $\beta_2$
are the response functions and $\pi$ is an undetermined
Lagrange multiplier which represents the reactive stress
arising by the incompressibility constraint.

The electric part of the stress (Maxwell stress) can be expressed by (see again \cite{BDO})
\begin{equation}
\begin{array}{l}
\T^{\tiny \mbox{\it M}}=\e (\bE \otimes \bE-\frac{1}{2} (\bE \cdot \bE) \I ),\vspace{0.2 cm}\\
T^{\tiny \mbox{\it M}}_{ij}=\displaystyle \e(E_{i}E_{j}- \frac{E^2}{2}\delta_{ij}),
\end{array}\label{def}
\end{equation}
where $E=|\bE|$. With these positions and without loss of
generality, the total current stress in an incompressible,
isotropic elastic and dielectrically homogeneous body can be
expressed as
\begin{equation}
\begin{array}{ccl}\label{CauchyGEN}
\T & = &
-p\,\I+\beta_1(\lambda_1,\lambda_2)\B+\beta_2(\lambda_1,\lambda_2)\B^{-1}+\e\bE\otimes\bE\\
T_{ij} & = &
-p\,\d_{ij}+\beta_1(\lambda_1,\lambda_2)B_{ij}+\beta_2(\lambda_1,\lambda_2)B_{ij}^{-1}+\e
E_i E_j,
\end{array}
\end{equation}
having set $p=\displaystyle \pi+\frac{\e}{2}E^2$. Thus, the principal stresses have the values
\begin{equation}\label{Cauchycomponents}
t_i=T_{ii}=-p+\beta_1(\lambda_1,\lambda_2)\lambda_i^2+\beta_2(\lambda_1,\lambda_2)\lambda_i^{-2}+\e
E_i^2.
\end{equation}

\section{Tensile stretches region}

Consider a thin elastic sheet which is made of isotropic,
incompressible material, whose upper and lower faces
are bonded to compliant electrodes. The reference configuration
is a stress free state with zero
applied voltage; we here assume that this configuration coincides
with a right cylindrical region with flat mid-surface $\Omega$ and
constant thickness $h$. Under the assumption that $h$ is small as compared with $\Omega$
`diameter', we embrace the
\emph{membrane approximation} which asserts that the bending
stiffness is zero and that any in-plane compressive stress
immediately leads to the membrane buckling, with the appearance of
wrinkled regions.

According with most common application schemes of EAPs we assume that $\Omega$ remains flat
after deformation. We also assume
that orthogonal fibers to the plane of $\Omega$
remain orthogonal to this plane also after deformation. We consider thickness variations that, by the
incompressibility hypothesis, are accomplished by
compatible variations of the in-plane stretches, so that \begin{equation}\label{incompr}
\lambda_3=\frac{1}{\lambda_1\lambda_2}
\end{equation}
where $\be_3$ is the unit vector orthogonal to $\Omega$.

The application of a voltage on the electrodes
determines the insurgence of an electric field $\bE$ which should
be rigorously calculated by solving the corresponding electromechanical equilibrium problem  (see e.g. \cite{BDO}).
On the other hand, since each electrode is an equipotential surface, coherently with the assumption of
preservation of the direction of normal fibers and of membrane thinness,
we assume that the electric field remains
perpendicular to $\Omega$. Of course this hypothesis
fails at the boundary of the membrane and in correspondence
with possible deformation localization, but typically, with the hypothesis
of small thickness, this approximation can be energetically justified
(e.g. \cite{Par}).
Under the described assumptions, if a voltage $V$ is applied to the electrodes,
then the electric field at any point of the current configuration
amounts to
\begin{equation}\label{E}
\bE=\frac{V}{h\lambda_3}\be_3.
\end{equation}

While the approach that we consider in the following is general,
to fix the ideas, we here consider a diffuse constitutive assumption for polymeric materials,
i.e. the Mooney-Rivlin
constitutive model, characterized by constant response functions:
\begin{equation}\label{constMR}
\beta_1 = 2c_1,\hspace{10pt}\beta_2=-2c_2,
\end{equation}
with $c_1\geq 0$ and $c_2\geq 0$.
It is easy to check that for
this material class the shear modulus is given by
$\mu=2(c_1+c_2)$, which means that stiffer materials are endowed
of higher values of the constants $c_1$ and $c_2$.

Under these hypotheses  (\ref{Cauchycomponents}) gives
\begin{equation}\label{CauchyMRcomponents}
\begin{array}{ccl}
t_1 & = & -p+2c_1\lambda_1^2-2c_2\lambda_1^{-2}\vspace{0.2 cm}\\
t_2 & = & -p+2c_1\lambda_2^2-2c_2\lambda_2^{-2}\vspace{0.2 cm}\\
t_3 & = & -p+2c_1\lambda_1^{-2}\lambda_2^{-2}+2(k_{\mbox{\tiny{\it V}}}-c_2)\lambda_1^2\lambda_2^2.\\
\end{array}
\end{equation}
Here, as proposed in \cite{DPSZ}, we introduce
\begin{equation}\label{kV}
k_{\mbox{\tiny{\it V}}} = \frac{\e V^2}{2 h^2}
\end{equation}
measuring the electric energy density and representing our activation parameter.

The undetermined multiplier $p$ can be deduced by imposing the
boundary condition $t_3=0$ on the upper and lower faces:
\begin{equation}\label{p}
p=2c_1\lambda_1^{-2}\lambda_2^{-2}+2(k_{\mbox{\tiny{\it V}}}-c_2)\lambda_1^2\lambda_2^2.
\end{equation}
After substitution in Eq.s (\ref{CauchyMRcomponents}), the
in-plane principal stresses
are given by
\begin{equation}\label{stess2D}
\begin{array}{ccl}
t_1 &=&
2[c_1(\lambda_1^2-\lambda_1^{-2}\lambda_2^{-2})-c_2(\lambda_1^{-2}-\lambda_1^2\lambda_2^2)-k_{\mbox{\tiny{\it V}}}\lambda_1^2\lambda_2^2]\vspace{0.2 cm}\\
t_2 &=&
2[c_1(\lambda_2^2-\lambda_1^{-2}\lambda_2^{-2})-c_2(\lambda_2^{-2}-\lambda_1^2\lambda_2^2)-k_{\mbox{\tiny{\it V}}}\lambda_1^2\lambda_2^2].
\end{array}
\end{equation}

\begin{figure}[htb]
\begin{centering}\vspace{-2 cm}
\includegraphics[width=12cm]{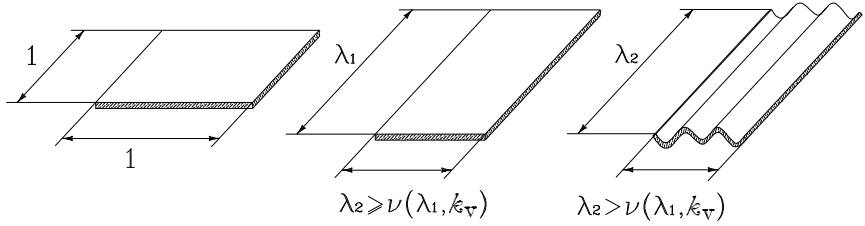}
\par\end{centering}\vspace{-3 cm}
\caption{\label{Fig1} \setlength{\baselineskip}{10 pt}{\footnotesize
Scheme of the natural strain and membrane wrinkling.}}
\end{figure}

We are now in position to introduce the central idea of
\emph{natural width in simple tension}, first formulated by Pipkin
in his seminal work \cite{Pipkin1} and later developed within
the context of \emph{Tension Field Theory} of thin elastic
membranes (e.g., among many others, \cite{SteigPip},
\cite{Steig1}).

Consider a state of local uniaxial stress in direction
(say) $\be_1$; under the assumption $t_2=t_3=0$, the transverse stretch in direction
$\be_2\perp\be_1$ assumes a specific value called \emph{natural
width} in tension, which is constitutively dependent on the
stretch $\lambda_1$ in direction $\be_1$
\begin{equation}\label{naturalwidth}
\lambda_2=\nu(\lambda_1, k_v).
\end{equation}
Since for
$\lambda_2=\nu(\lambda_1, k_v)$ it is $t_2=0$, any attempt to reduce the transverse stretch under this value requires a the application of a compressive stress, leading to the formation of wrinkles
(see Fig.\ref{Fig1}). This hypothesis of the tension field theory (see \cite{Steig1}) on the material behavior can be easily shown to hold in the case here considered of Mooney-Rivlin
materials.

While in the classical tension field theory, without electric field, it results
$\lambda_2=\nu (\lambda_1)=\lambda_1^{-1/2}$,
in the present case the natural width depends on the applied voltage and in view of Eq. (\ref{stess2D})$_2$ takes the form
\begin{equation}\label{naturalwidth1}
\lambda_2=\nu(\lambda_1,k_v)=\lambda_1^{-1/2}\left[\frac{c_1+c_2\lambda_1^2}
{c_1+c_2\lambda_1^2-k_{\mbox{\tiny{\it V}}}\lambda_1^2}\right]^{1/4}.
\end{equation}
Analogous considerations hold for uniaxial tension in direction
$\be_2$, so that the condition $t_1=0$ gives the natural width in
the transverse direction $\be_1$
\begin{equation}\label{naturalwidth2}
\lambda_1=\nu(\lambda_2,k_v)=\lambda_2^{-1/2}\left[\frac{c_1+c_2\lambda_2^2}
{c_1+c_2\lambda_2^2-k_{\mbox{\tiny{\it V}}}\lambda_2^2}\right]^{1/4}.
\end{equation}

As a consequence we have that:
{\it for any given voltage $V$, the membrane is in traction
when $\lambda_1>\nu(\lambda_2,k_v)$ and
$\lambda_2>\nu(\lambda_1,k_v)$. In all other cases, the membrane undergoes a compression-induced instability.}

\begin{figure}[h!]
\begin{centering}
\includegraphics[width=12cm]{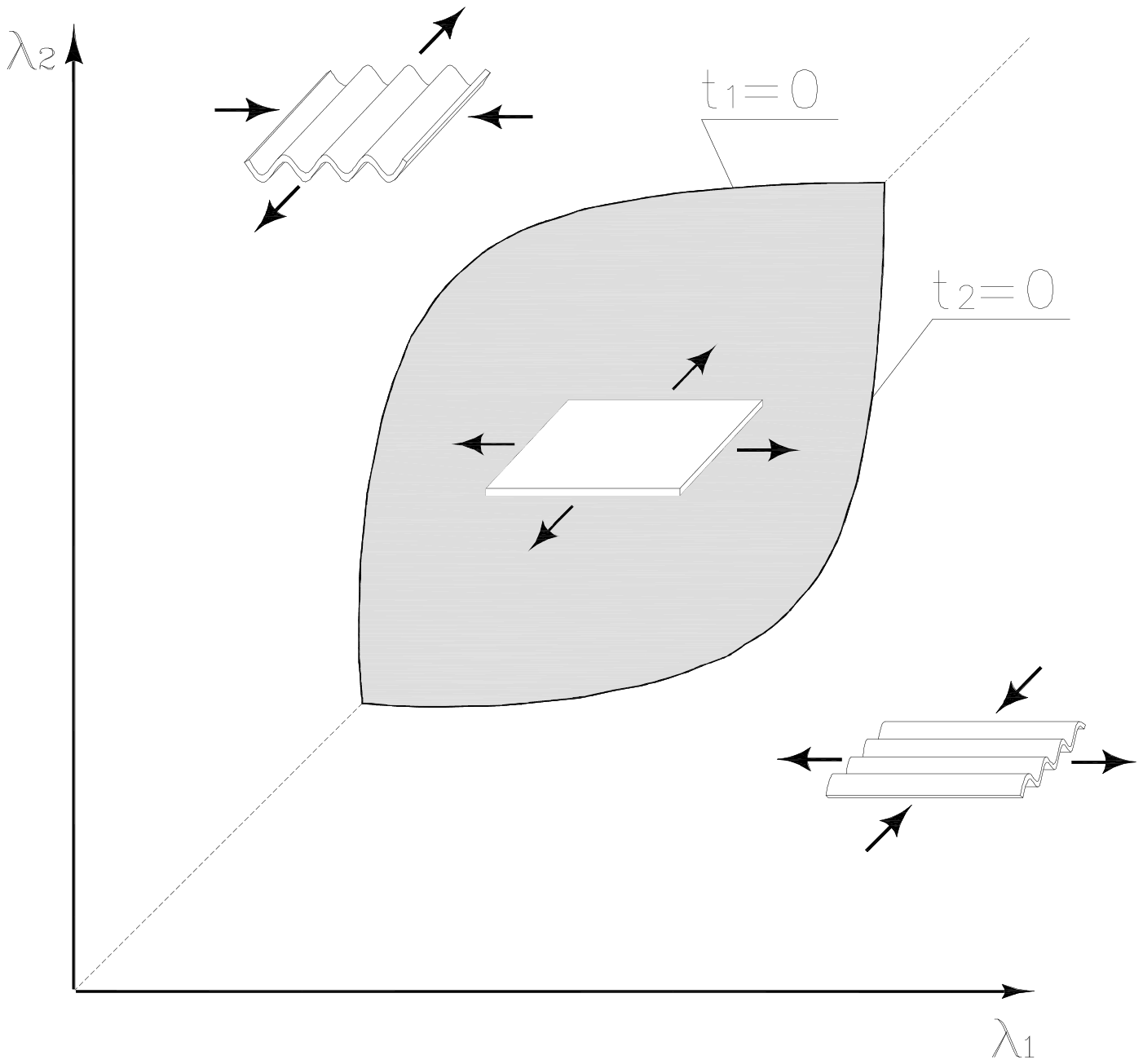}
\vspace{-7.25 cm}
$$\hspace{2 cm}\scrD$$
\vspace{4.75 cm}
\par\end{centering}
\caption{\label{Fig2}\setlength{\baselineskip}{10 pt}{\footnotesize
Region $\scrD$ of tensile states. }}
\end{figure}

In other words we deduce the existence of a voltage-dependent
region in the principal stretches space (see Fig.\ref{Fig2})
\begin{equation}\label{D}\scrD (k_v)
=\left\{(\lambda_1,\lambda_2):\lambda_1>\nu(\lambda_2,k_v),\hspace{5pt}
\lambda_2>\nu(\lambda_1, k_v)\right\}
\end{equation}
that collects the
possible values of $(\lambda_1,\lambda_2)$ corresponding to tensile
states. Wrinkling arises for combinations of the principal
stretches which do not belong to $\scrD$. The two boundaries of $\scrD$ represent the
states with
$t_1=0$ or $t_2=0$, whereas the two vertexes represent the equibiaxial configurations
with $t_1=t_2=0$. 

\begin{figure}[htb]
\begin{centering}
\includegraphics[width=10 cm]{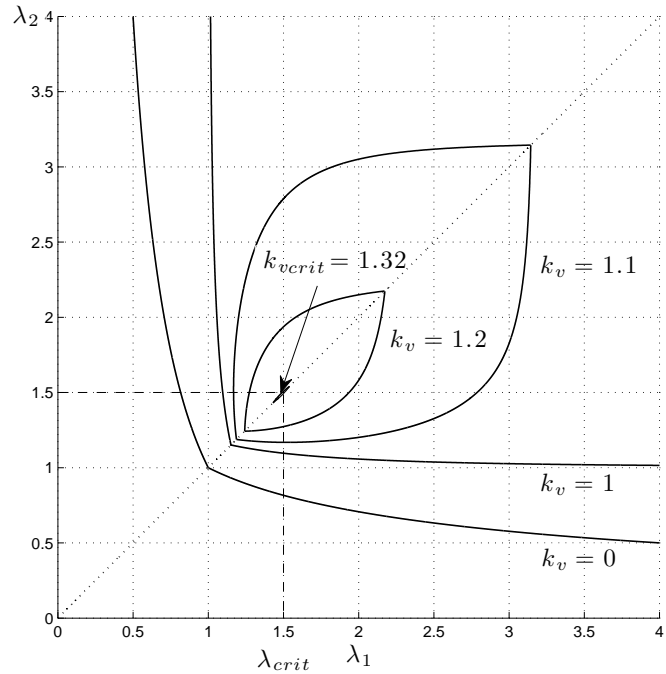}
\par\end{centering}
\caption{\label{Fig3}\setlength{\baselineskip}{10 pt}{\footnotesize
Dependence of the tensile region $\scrD$ on the activation parameter $k_v$.
Here  for $c_1=1$ and $c_2=1$.
The critical values
$k_v^{crit}$ and $\lambda^{crit}$ are the activtion thresholds leading to the disappearance
of tensile states.}}
\end{figure}

Observe that the two curves of the boundary are symmetric
with respect to the line
$\lambda_1=\lambda_2$. Thus in the following we restrict our attention to the curve $t_2=0$
 for $\lambda_1>\lambda_2$.
When the applied voltage is zero, then
$\nu(\lambda_1,0)=\lambda_1^{-1/2}$. In this case, the domain $\scrD$
is unbounded. The application of a voltage modifies the domain
$\scrD$ as follows (see Fig.\ref{Fig3}). Since $c_1\geq 0$ and $c_2\geq
0$, a straightforward analysis shows that for
$k_{\mbox{\tiny{\it V}}}<c_2$ the domain $\scrD$ remains unbounded, whereas the boundary edges are shifted away from the origin.
As soon as the voltage overcomes the threshold $k_{\mbox{\tiny{\it
V}}}=c_2$,  the function $\nu$ has a vertical
asymptote in correspondence to the stretch
\begin{equation}\label{asympt}
\lambda_1=\lambda^*=\sqrt{\frac{c_1}{k_{\mbox{\tiny{\it V}}}-c_2}}.
\end{equation}
A simple analysis now reveals that for lower values of the
voltage parameter $k_{\mbox{\tiny{\it V}}}$ the two symmetric boundary curves of the domain $\scrD$  intersect at the upper and lower vertexes corresponding to stretches
 $\lambda_1=\lambda_2>1$ and
$\lambda_1=\lambda_2<\lambda^*$, respectively.

As we show in Fig.\ref{Fig3}, by increasing the activation parameter
$k_{\mbox{\tiny{\it V}}}$ the two vertexes approach each other
until they coalesce and no configuration is tensile.
\begin{figure}[h!]\vspace{-2 cm}\begin{centering}
\includegraphics[width=10 cm]{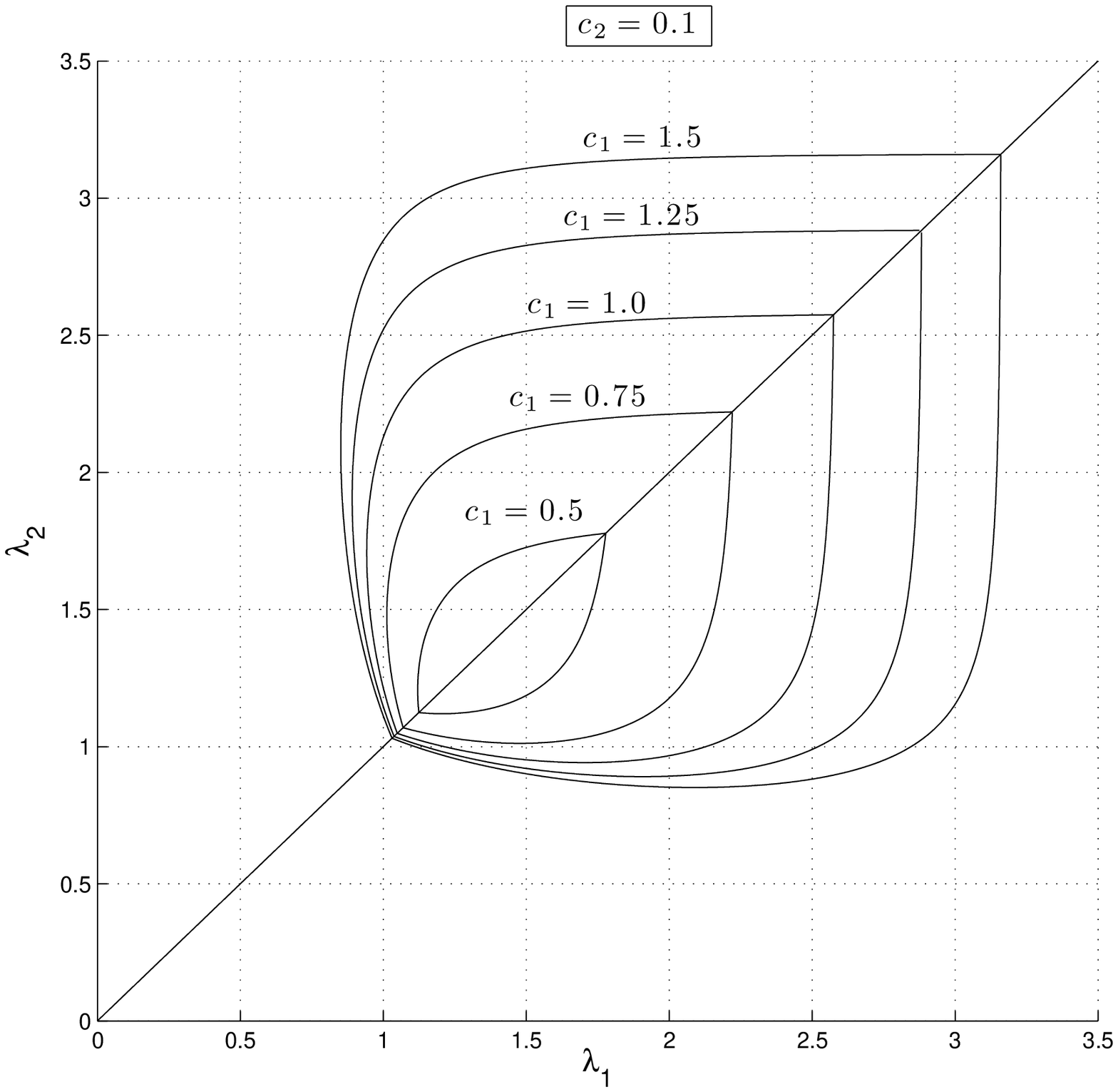} a)\vspace{-0.4 cm}
\includegraphics[width=10 cm]{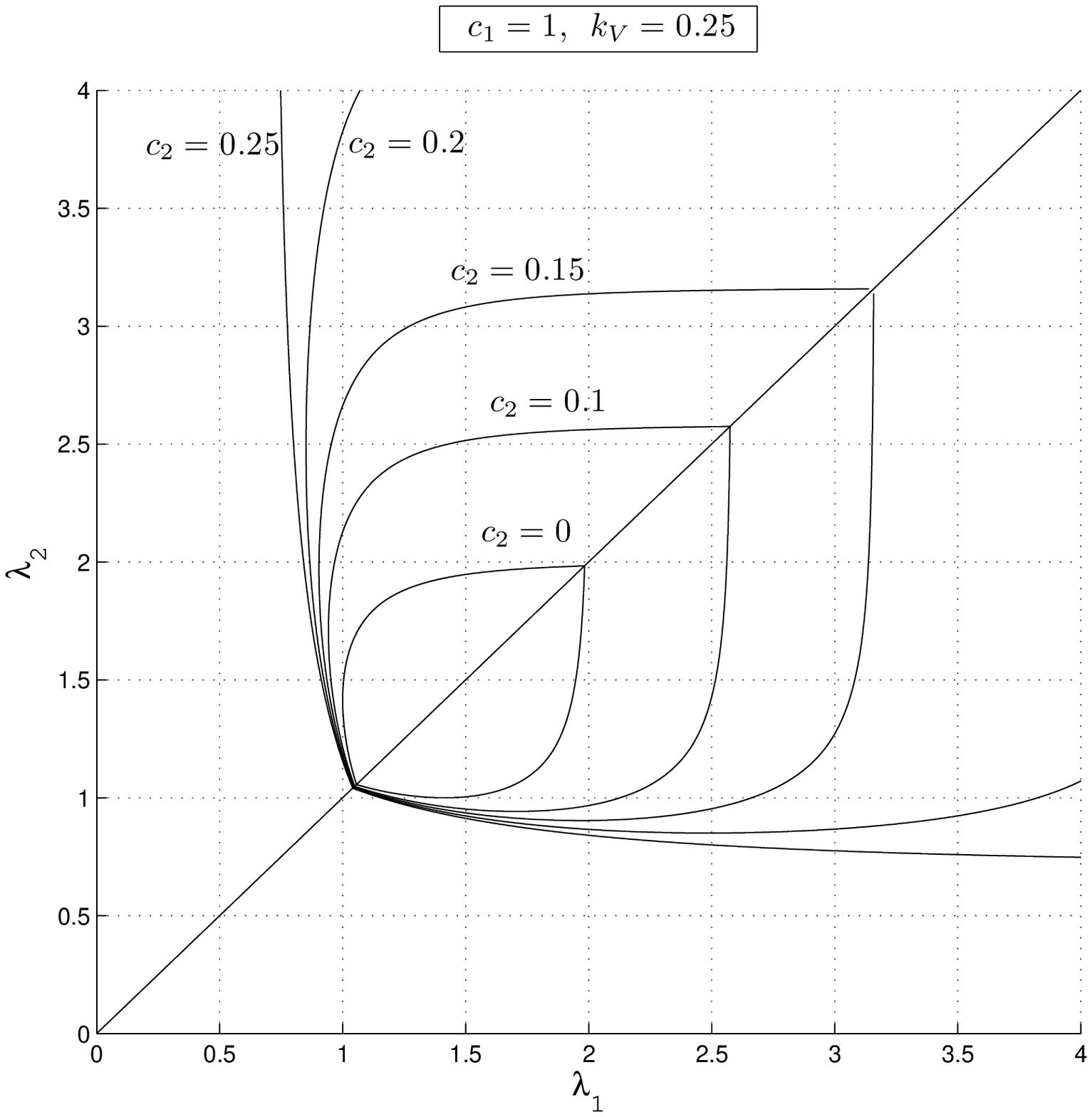}b)\vspace{-0.15 cm}
\par\end{centering}
\caption{\setlength{\baselineskip}{10 pt}{\label{Fig4}
\footnotesize
Dependence of the tensile region $\scrD$
on the constitutive parameters $c_1$ and $c_2$.}}
\end{figure}
We thus
deduce the existence of a critical threshold $k_{\mbox{\tiny{\it
V}}}=k_{\mbox{\tiny{\it V}}}^{crit}$, such that for
$k_{\mbox{\tiny{\it V}}}>k_{\mbox{\tiny{\it V}}}^{crit}$ there is
no stable equilibrium configuration. We call $\lambda^{crit}$
the corresponding stretch threshold (see again Fig.\ref{Fig3}).

We point out that a similar approach can be extended to the more
general case of non constant response functions
$\beta_1=\beta_1(\lambda_1,\lambda_2)$ and $\beta_2=\beta_2(\lambda_1,\lambda_2)$ with the boundaries of $\scrD$ obtained
by numerically solving $t_1=0$ and $t_2=0$ in (\ref{stess2D}).
Moreover, we remark that in this analysis we consider only
compression induced instabilities, but other types of purely
mechanical or electromechanical instabilities can be important
(see \cite{GP}, \cite{ZHS} and \cite{ZH} and references therein).

Finally, in Fig.\ref{Fig4} we show the dependence of the tensile region, for a fixed value of $k_v$, on the constitutive parameter $c_1$ and $c_2$. Observe that in both cases the stiffer is the material, the wider is the region of tensile stretches configurations. It is important to observe that the proposed approach provides an immediate tool for the study of the EAPs behavior regarding compression instability and thus it may reveal its importance in the field of material design.

\section{Two simple applications}

In this section, as illustrative example, we apply our analysis
to homogeneously deformed EAP sheets under different boundary
conditions. Despite our approach is not limited
by specific constitutive assumptions or boundary conditions,
we here take into consideration some
simple cases which are amenable of fully analytic solutions and
allow an easy interpretation of the results.

Consider firstly the case of Neo-Hookean materials, i.e.
$c_2=0, c_1=\mu/2$, where $\mu$ is the shear modulus. Hence
(\ref{naturalwidth1}) gives
\begin{equation}\label{naturalwidth1NH}
\lambda_2=\nu(\lambda_1, k_{\mbox{\tiny{\it V}}})=\lambda_1^{-1/2}\left[\frac{\mu}
{\mu-2 k_{\mbox{\tiny{\it V}}}\lambda_1^2}\right]^{1/4}.
\end{equation}

\noindent In this case the two  vertexes
of the region $\scrD$,  with $\lambda_1=\lambda_2=\lambda$, are the solutions of
\begin{equation} \label{equi}2 k_v \lambda^8 - \mu \lambda^6 + \mu=0. \end{equation}

\noindent These vertexes coalesce for
\begin{equation}\label{kcrit}k_{\mbox{\tiny{\it
V}}}=k_{\mbox{\tiny{\it V}}}^{crit}=\frac{3\mu}{
2^{\frac{11}{3}}}\end{equation}

\noindent which corresponds to an
equibiaxial strain
\begin{equation}\label{lambdacrit}
\lambda_1=\lambda_2=\lambda^{crit}=2^{\frac{1}{3}}.
\end{equation}
It should be remarked that the simplicity of
(\ref{kcrit})  and (\ref{lambdacrit}) is due to the Neo-Hookean constitutive assumption, which is satisfactory only at low stretches.
For large stretches the entropic hardening effect, which is not accounted by the Neo-Hookean law, can play an important role in modifying the $\scrD$ region (see e.g.
Fig.\ref{Fig4}).

We consider now the two following cases, respectively without and with assigned prestretch.

\subsubsection*{The case without prestretch}

\begin{figure}[htb]\begin{centering}
\includegraphics[width=15 cm]{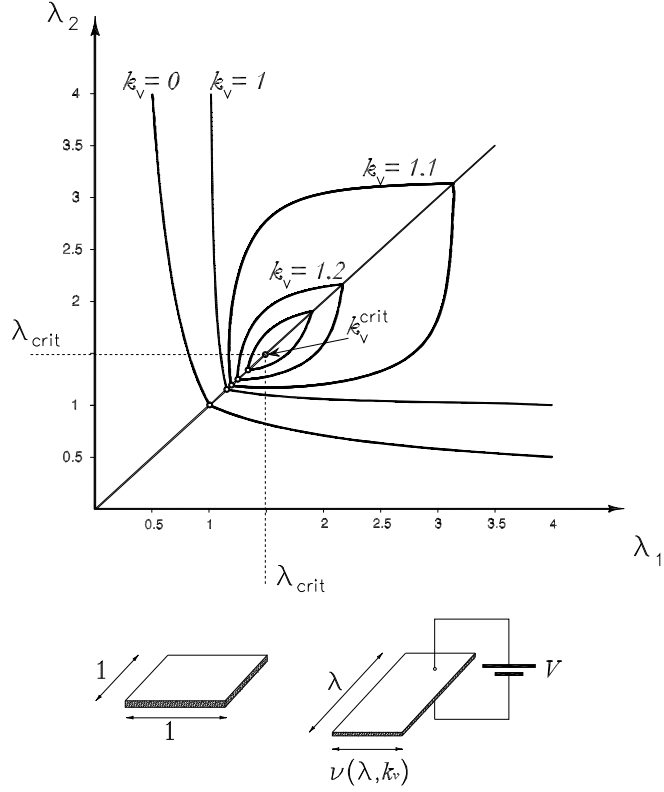}
\par\end{centering}
\caption{\label{Fig5}
\setlength{\baselineskip}{10 pt}{\footnotesize
Equilibrium solutions under the hypothesis of homogeneous deformation, in the non prestretched case. Circles represent the equilibrium states for the different values of the activation parameter $k_v$.}}
\end{figure}

Consider first the case of  an EAP membrane under an assigned voltage $V$ (see the scheme in Fig.\ref{Fig5})
and no prestretch at the boundaries. By imposing that $t_1=t_2=t_3=0$ we obtain that the equilibrium solutions correspond to the
intersection of $\scrD$ with the line $\lambda_1=\lambda_2=\lambda$.
Thus we are in the case of equibiaxial strain with the in-plane stretch $\lambda$ satisfying Eq. (\ref{equi}).
As a consequence we may interpret the vertexes of $\scrD$ as the stretches corresponding to the present situation. Observe that for given activation $k_{\mbox{\tiny{\it V}}}$ there are two equilibrium solutions.  Moreover, the stretch of the equilibrium solution corresponding to the upper vertex decreases as $k_{\mbox{\tiny{\it V}}}$ grows. Thus the thickness of the membrane grows with $k_{\mbox{\tiny{\it V}}}$ so that we may argue that this equilibrium solution is unstable. This is in accordance with the results in \cite{DPSZ} where two equilibrium solutions have been obtained for each value of the activation parameter and the larger equilibrium stretch corresponds to an unstable state.

Based on previous analysis, we may deduce that when we increase $k_{\mbox{\tiny{\it V}}}$ there exists a critical value of $k_{\mbox{\tiny{\it V}}}=k_{\mbox{\tiny{\it V}}}^{crit}$
for which the tensile region disappears. This maximum activation value grows with the stiffness of the material. The corresponding limit activation in-plane stretch is given in (\ref{lambdacrit}).
After this threshold no equilibrium solution is possible. This effect represents what is called in the literature as pull-in instability (see \cite{DPSZ}).

\subsubsection*{The prestretched case}

\begin{figure}[htb]
\begin{centering}
\includegraphics[width=15 cm]{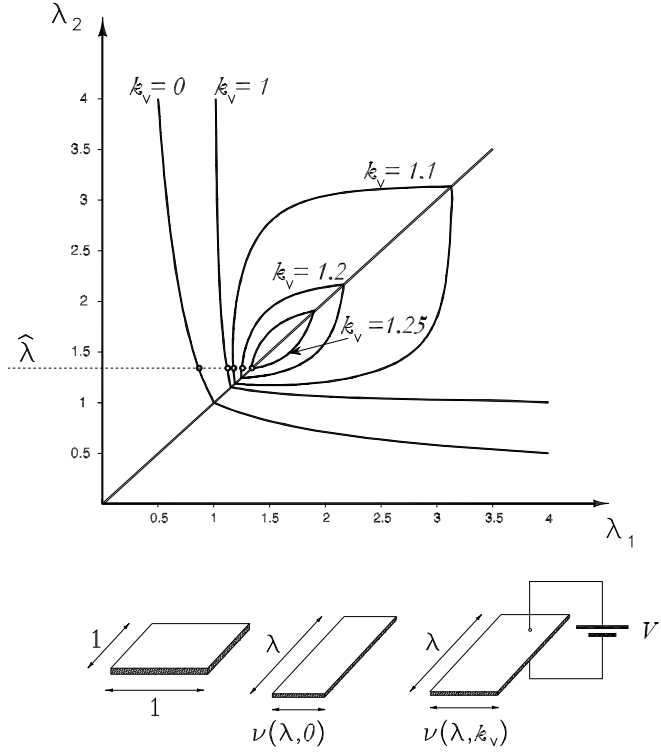}
\par\end{centering}
\caption{\label{Fig6} \setlength{\baselineskip}{10 pt}{\footnotesize
Equilibrium solutions under the hypothesis of homogeneous deformation, in the prestretched case. Circles represent the equilibrium states for the different values of the activation parameter $k_v$. }}
\end{figure}

Consider now the hypothesis that a prestretch $\lambda_2=\hat \lambda_2$ is assigned in direction (say) $\be_2$ of a rectangular EAP membrane (see Fig.\ref{Fig6}).
The homogeneous equilibrium solution is obtained by requiring $t_1=t_3=0$. For given  $k_{\mbox{\tiny{\it V}}}$  and $\hat\lambda_2$, the stretch $\lambda_1$  is given by (\ref{naturalwidth1}) as $\lambda_1=\nu(\hat \lambda_2,  k_{\mbox{\tiny{\it V}}}$ ).
Then we may interpret the boundary of $\scrD$, i.e. the curves of the natural widths, as representing the equilibrium solutions in the prestretched case.

Observe that the system looses its equilibrium for an activation $k_{\mbox{\tiny{\it V}}}$ (see Fig.\ref{Fig6})  such that the straight line $\lambda_1=\hat \lambda_1$ corresponds to one of the two vertexes of the tensile region. Thus, the largest activation $ k_{\mbox{\tiny{\it V}}}=k_{\mbox{\tiny{\it V}}}^{crit}$
if one chooses a prestretch $\hat \lambda_2=\lambda^{crit}$. We recall that the existence of an optimal prestretch is also experimentally deduced in \cite{PKP, SGS} and theoretically described in \cite{DPSZ} .

\newpage

\vspace{1 cm}
\noindent {\bf Acknowledgment}. The authors have been supported by MIUR project PRIN $2008$, `Modelli multiscala per strutture in materiali innovativi'
and the project, Progetto di ricerca industriale-Regione Puglia, `Modelli innovativi per sistemi meccatronici'.





\bibliographystyle{model1a-num-names}



\end{document}